\documentclass[proof]{pasj01}
\Received{2022/10/03}
\Accepted{2022/11/07}
\onecolumn
\usepackage{natbib}
\usepackage{color}
\usepackage{bm}
\usepackage{lineno}
\usepackage[normalem]{ulem}
\graphicspath{{./}{figures/}}
\newcommand{\omegas}{\omega_{\star}}
\newcommand{\thetao}{\theta_{\rm o}}
\newcommand{\thetaS}{\theta_{\star}}
\newcommand{\dts}{\Delta\theta_{\rm s}}
\newcommand{\ILD}{I_{\rm LD}}
\newcommand{\dI}{\Delta I}
\newcommand{\varphic}{\varphi_{\rm c}}
\newcommand{\varphics}{\varphi_{\rm c,s}}

\begin{document}
\title{Modeling photometric variations due to a global
  inhomogeneity on an obliquely rotating star: application to
  lightcurves of white dwarfs}
\author{
  Yasushi \textsc{Suto}\altaffilmark{1,2,3}, 
  Shin \textsc{Sasaki}\altaffilmark{4},
  Masataka \textsc{Aizawa}\altaffilmark{5}, 
  Kotaro \textsc{Fujisawa}\altaffilmark{1}, and
  Kazumi \textsc{Kashiyama}\altaffilmark{6,7}}
\email{suto@phys.s.u-tokyo.ac.jp}
\altaffiltext{1}{Department of Physics, The University of Tokyo, Tokyo
113-0033, Japan}
\altaffiltext{2}{Research Center for the Early Universe, School of
Science, The University of Tokyo, Tokyo 113-0033, Japan}
\altaffiltext{3}{Laboratory of Physics, Kochi University of Technology,
  Tosa Yamada, Kochi 782-8502, Japan}
\altaffiltext{4}{Department of Physics, Tokyo Metropolitan University,
Hachioji, Tokyo 192-0397, Japan}
\altaffiltext{5}{Tsung-Dao Lee Institute, Shanghai Jiao Tong University, Shengrong Road 520, 201210 Shanghai, P. R. China}
\altaffiltext{6}{Astronomical Institute, Tohoku University,
Sendai 980-8578, Japan}
\altaffiltext{7}{Kavli Institute for the Physics and Mathematics
  of the Universe, The University of Tokyo, Kashiwa 277-8583, Japan}

\maketitle

\begin{abstract}
  We develop a general framework to compute photometric variations
  induced by the oblique rotation of a star with an axisymmetric
  inhomogeneous surface.  We apply the framework to compute
  lightcurves of white dwarfs adopting two simple models of their
  surface inhomogeneity. Depending on the surface model and the
  location of the observer, the resulting lightcurve exhibits a
  departure from a purely sinusoidal curve that are observed for a
  fraction of white dwarfs. As a specific example, we fit our model to
  the observed phase-folded lightcurve of a fast-spinning white dwarf
  ZTF J190132.9+145808.7 (with the rotation period of 419s). We find
  that the size and obliquity angle of the spot responsible for the
  photometric variation are $\dts \approx 60^\circ$ and $\thetaS
  \approx 60^\circ$ or $90^\circ$, respectively, implying an
  interesting constraint on the surface distribution of the magnetic
  field on white dwarfs.
\end{abstract}


\section{Introduction}

A number of white dwarfs (WDs) have exhibited periodic photometric
variations~\citep[e.g.,][]{Achilleos1992,Barstow1995,Wade2003,Brinkworth2005,Brinkworth2013,Reding2020,Caiazzo2021,Kilic2021,Williams2022}.
Those variations are commonly interpreted to originate from
inhomogeneities on the stellar surface that obliquely rotate around
the spin axis. Those inhomogeneities may be produced by magnetic spots
in a convective atmosphere~\citep[e.g.,][]{Valyavin2008,Valyavin2011}
or the so-called magnetic dichroism~\citep{Ferrario1997}; the
continuum opacity changes with rotational phase according to the
amplitude of the magnetic field strength across the stellar disk.  As
a result, the photometric flux varies as well.  Hence, photometric
lightcurves of WDs carry rich information on the distribution of the
surface magnetic fields, which may be further disentangled by
combining the spectro-polarimetric data if available
\citep[e.g.,][]{Liebert1977,Donati1994,Euchner2002,Valyavin2014}.

For instance, \citet{Brinkworth2013} have performed time-series
photometry of 30 isolated magnetic WDs, and 5 (24 \%) are variable
with reliably measured periods of (1 -- 7) hours.  Quite
interestingly, all of their lightcurves are well fitted by a
monochromatic sinusoidal curve. Indeed, apart from 9 WDs whose
lightcurves are significantly contaminated by their varying comparison
stars, they find that 14 out of the remaining 21 show evidence for
variability whose lightcurves are consistent with a monochromatic
sinusoidal curve. In marked contrast to other stars for which
non-sinusoidal rotation signatures are commonly observed
\citep[e.g.,][]{2013ApJ...767...60R}, the sinusoidal lightcurve seems
to be fairly generic for these types of WDs, which have a relatively
stable surface structure over time.

Thanks to the Kepler satellite, Transiting Exoplanet Survey Satellite
(TESS), and Zwicky Transient Facility (ZTF)
\citep[][]{Maoz2015,Reding2020,Hermes2021} among others, the number of
photometric lightcurves for WDs with a high accuracy and cadence is
rapidly increasing. Indeed, fast-spinning WDs with a rotation period
even less than 10 minutes have been recently
discovered~\citep[][]{Reding2020,Caiazzo2021,Kilic2021}.  For such
cases, however, it may be due to their pulsation, instead of the
rotation, since their frequency ranges are
overlapped~\citep[e.g.,][]{2008ARA&A..46..157W,2020NatAs...4..663H}.
This points to an importance of quantitative modeling of photometric
lightcurves due to inhomogeneities on rotating stars.  For instance,
the phase-folded lightcurve of one of the fast-spinning WDs (ZTF
J190132.9+145808.7) seems to exhibit a non-sinusoidal
feature~\citep[][and see Sec. \ref{subsec:application2ZTF}
  below]{Caiazzo2021}, which could be relatively common for these
young massive WDs.  Thus, such modeling of photometric variations due
to an inhomogeneity on the stellar surface should become even more
crucial in the coming era of the Rubin Observatory LSST
Camera~\citep{2019ApJ...873..111I}.

Photometric rotational variations due to multiple circular starspots
have been studied previously
\citep[e.g.,][]{Budding1977,Dorren1987,Eker1994,Landolfi1997,Kipping2012}.
More recently, \citet{SSNB2022} proposed a fully analytic model for
the photometric variations due to infinitesimally small multiple
starspots on a differentially rotating star, and performed a series of
mock Lomb-Scargle analysis relevant to the Kepler data
\citep[e.g.,][]{LBKS2022}.  In this paper, instead, we focus on an
inhomogeneous surface intensity pattern that is axisymmetric around an
axis misaligned to the stellar spin axis. The inhomogeneity produces
periodic photometric variations due to the stellar rotation.  The
purpose of the present paper is to develop a general formulation to
describe the rotational variation, compute the resulting lightcurves
for a couple of physically-motivated specific models for WDs, and
apply the methodology to derive constraints on several parameters for
ZTF J190132.9+145808.7 as an example.

The rest of the paper is organized as follows.  Section
\ref{sec:formulation} presents our formulation of the lightcurve
modeling for a solid-body rotating star with an inhomogeneous surface
intensity distribution. We first present a general formulation to
compute the photometric modulation due to the oblique stellar rotation
by adopting a quadratic limb darkening law. While our formulation
would be essentially the same as those in the previous literature, the
resulting expressions are characterized by a set of parameters with
clear physical interpretations, and thus more useful in fitting to the
observed lightcurves. In order to show the advantage of our method, we
apply the formulation to two simple inhomogeneity models in section
\ref{sec:application}: a constant-intensity single circular spot
(cap-model), and a globally varying-intensity surface (p-model). We
find that the p-model leads to a sinusoidal lightcurve strictly, while
the cap-model exhibits a departure from a sinusoidal lightcurve
depending on the geometrical configurations between the starspot and
the observer. In both models, the photometric variations are fairly
insensitive to the limb darkening effect.  In subsection
\ref{subsec:application2ZTF}, we show that the cap-model well explains
the observed lightcurve of ZTF J190132.9+145808.7.
Finally, section \ref{sec:conclusion} is devoted to discussion and
conclusion of the present paper. Analytic derivations of several
integrals appearing in the main text are given in Appendix.

\section{Basic formulation of the lightcurve modeling for an obliquely
  rotating star
  \label{sec:formulation}}

\begin{figure}
 \begin{center}
 \includegraphics[width=8cm]{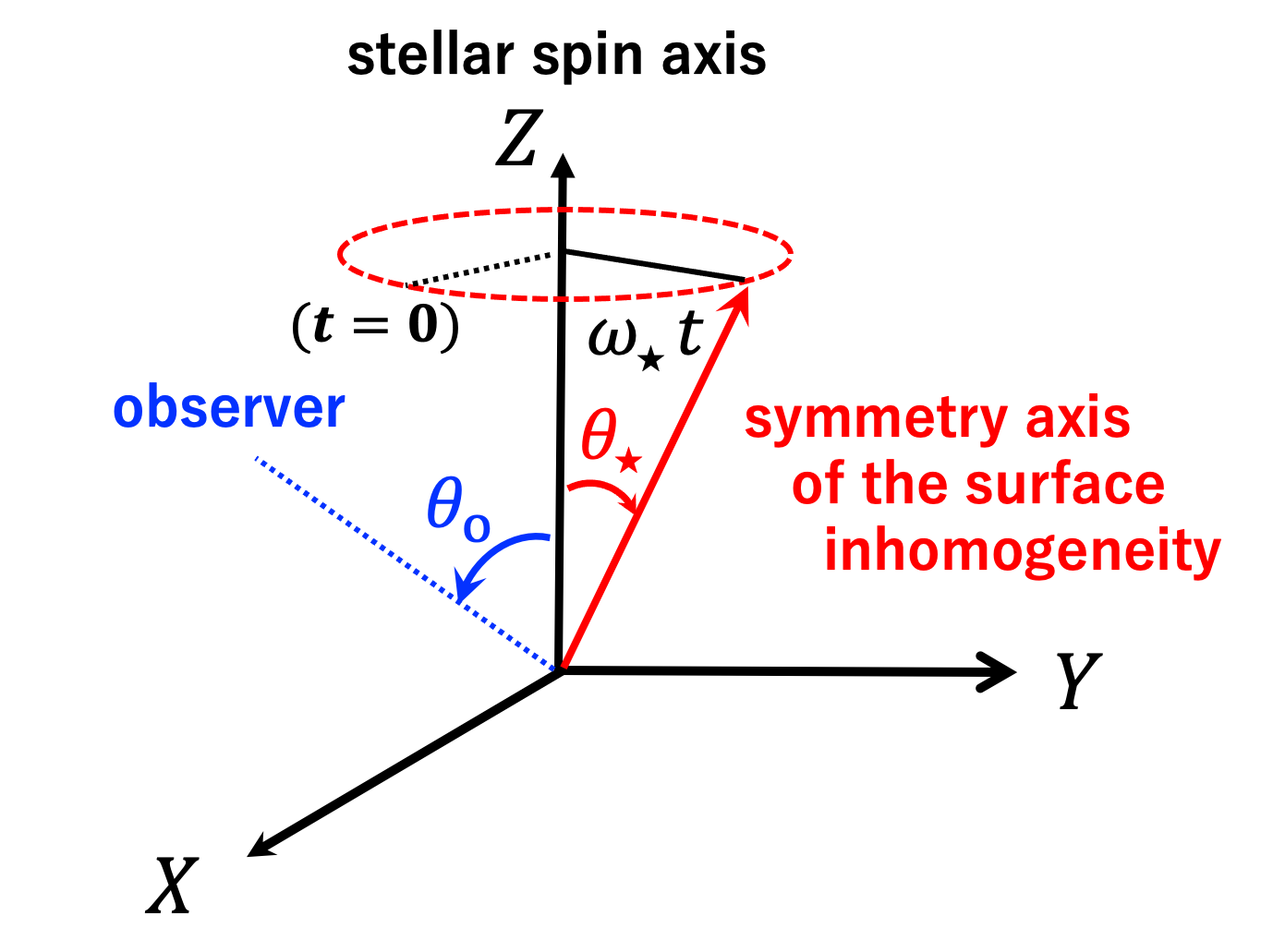} 
 \end{center}
 \caption{Schematic illustration of the star and the observer.  The
   observer is located at $(\sin \thetao, 0, \cos\thetao)$.  In this
   frame, the stellar spin axis is chosen as the $Z$-axis, and the
   unit vector of symmetry axis is defined to be $(\sin \thetaS, 0,
   \cos\thetaS)$ at $t=0$.}
 \label{fig:wd-config}
\end{figure}

As shown in Figure \ref{fig:wd-config}, we consider a star rotating
along the $z$-axis with a spin angular frequency of $\omegas$, and the
observer is located at $(\sin \thetao, 0, \cos\thetao)$.  If we denote
the obliquity angle between the stellar spin axis and the symmetry
axis of the surface inhomogeneity distribution by $\thetaS$, the angle
between the observer's line-of-sight and the symmetry axis,
$\gamma(t)$ is given by
\begin{equation}
  \label{eq:gamma}
  \cos\gamma(t) = \sin\thetaS \sin\thetao \cos\omegas t
  + \cos\thetaS \cos\thetao ,
\end{equation}
and thus $\gamma(t)$ varies between $|\thetaS-\thetao|$ and
$\thetaS+\thetao$ ($<\pi$). In practice, the symmetry axis may
correspond to the magnetic dipole axis of WDs, or to the central axis
of a single spherical spot on the stellar surface.
While we assume that $\thetaS$ is constant throughout
the present analysis, its possible time-dependence is easily
incorporated by substituting the specific function $\thetaS(t)$ in
equation (\ref{eq:gamma}) because the photometric variation in our
formulation is completely specified by $\gamma(t)$ alone.

Following \citet{SSNB2022}, the normalized photometric lightcurve
of the stellar surface is
\begin{eqnarray}
\label{eq:Lt}
L(t) &=& \frac{\displaystyle 
  \iint K(\theta, \varphi) I(\theta,\varphi)
  \ILD(\theta,\varphi) \sin\theta\,d\theta\, d\varphi}
{\displaystyle \iint K(\theta, \varphi) \overline{I}
  \sin\theta\,d\theta\, d\varphi}\cr
&=& \frac{1}{\pi \overline{I}}
\iint K(\theta, \varphi) I(\theta,\varphi)  \ILD(\theta,\varphi)
\sin\theta\,d\theta\, d\varphi,
\end{eqnarray}
where $K$ is the weighting kernel of the surface visible to the
observer, $I(\theta,\varphi)$ and $\ILD(\theta,\varphi)$ indicate the
surface intensity distribution and the limb darkening (the edge
  of the stellar disk is observed to be dimmer than its central
  part), respectively, and the integration is performed over the
entire stellar surface \citep[see
  also,][]{Fujii2010,Fujii2011,Farr2018,Haggard2018,Nakagawa2020}.  We
introduce a constant surface intensity, $\overline{I}$, just for
normalization, but it is not directly determined from observed data
and can be chosen arbitrarily.

For an isotropically emitting stellar surface, the weighting kernel
$K$ is equivalent to the visibility computed from the direction cosine
between the unit normal vector of the stellar surface ($\bm{e}_\star$)
and the unit vector toward the observer (${\bm e}_{\rm o}$).  Without
loss of generality, one can define a spherical coordinate system in
which the symmetry axis is instantaneously set to be $z$-axis.
In this case, one obtains
\begin{eqnarray}
\label{eq:mustar}
\bm{e}_\star \cdot \bm{e}_{\rm o} \equiv \mu_\star
= \sin\gamma(t) \sin\theta [\cos\varphi +\Gamma(t)],
\end{eqnarray}
where
\begin{eqnarray}
\label{eq:def-Gamma}
\Gamma(t) \equiv  \cot\gamma(t)\cot\theta.
\end{eqnarray}
The location $(\theta, \varphi)$ on the surface is visible to the
observer if $\mu_\star>0$. Therefore,
the weighting kernel is simply written as
\begin{eqnarray}
\label{eq:kernel}
 K(\theta, \varphi)
= \max ( \mu_\star, 0 )
= \sin\gamma(t) \sin\theta~
\max[\cos\varphi +\Gamma(t), 0].
\end{eqnarray}
If we adopt a quadratic limb darkening law, $\ILD$ in equation
(\ref{eq:Lt}) is specified by the two parameters $u_1$ and $u_2$ as
\begin{eqnarray}
  \ILD(\theta,\varphi)
  =1-u_1(1-\mu_\star)-u_2(1-\mu_\star)^2
  = (1-u_1-u_2) +(u_1+2u_2)\mu_\star - u_2\mu_\star^2.
\end{eqnarray}
For instance, $u_1 = 0.47$ and $u_2 = 0.23$ for the Sun at $550$nm,
and $u_1 = 0.05$ and $u_2 = 0.51$ for a typical white dwarf with
$T_{\rm eff}=10^4$K and log $g$=8.0 in the LSST g-band
\citep{Cox2000book,Gianninas2013}.

We focus on a case where the surface intensity distribution is
axisymmetric, {\it i.e.,} $I(\theta,\varphi)=I(\theta)$. Then,
equation (\ref{eq:Lt}) reduces to
\begin{equation}
\label{eq:Lt-012}
L(t) = L_0(t)+L_1(t)+L_2(t),
\end{equation}
where
\begin{eqnarray}
\label{eq:Lt-0}
L_0(t) = (1-u_1-u_2) \frac{\sin\gamma}{\pi \overline{I}}
  \iint \max[\cos\varphi + \Gamma(t), 0]~
\sin^2\theta \, I(\theta)\, d\theta d\varphi,
\end{eqnarray}
\begin{eqnarray}
\label{eq:Lt-1}
L_1(t) = (u_1+2u_2) \frac{\sin^2\gamma}{\pi \overline{I}}
  \iint \max[\cos\varphi + \Gamma(t), 0]~
\sin^3\theta \, [\cos\varphi + \Gamma(t)]I(\theta)\, d\theta d\varphi,
\end{eqnarray}
and
\begin{eqnarray}
\label{eq:Lt-2}
L_2(t) = -u_2 \frac{\sin^3\gamma}{\pi \overline{I}}
  \iint \max[(\cos\varphi + \Gamma(t))^3, 0]~
\sin^4\theta \, I(\theta)\, d\theta d\varphi .
\end{eqnarray}

\begin{figure}
 \begin{center}
 \includegraphics[width=8cm]{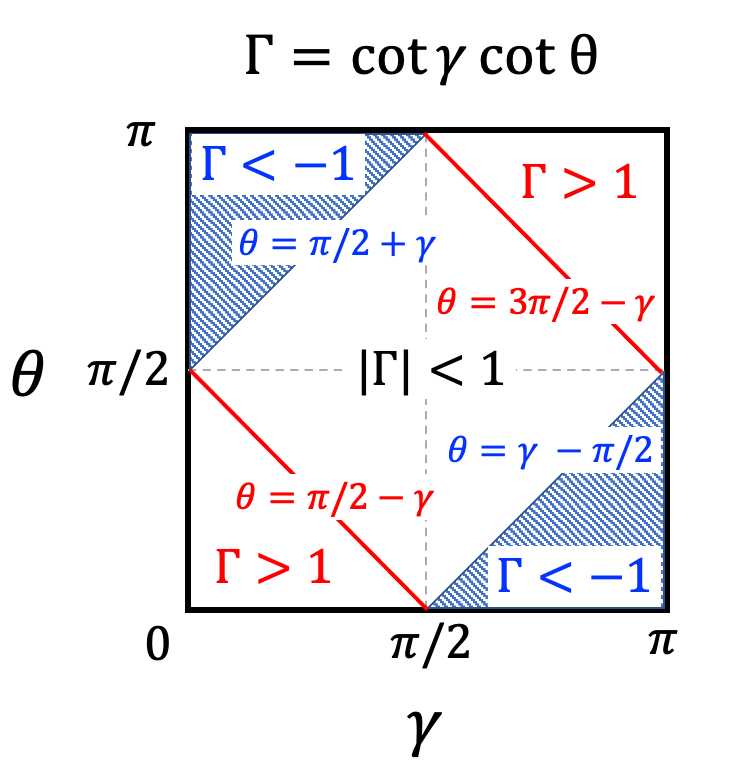} 
 \end{center}
 \caption{Range of $\theta$ and $\gamma$ for the visible part of the
   stellar sphere for the observer. The region with $\Gamma<-1$
   corresponds to the position $(\theta, \varphi)$ that is invisible
   to the observer, while that with $\Gamma>1$ is always visible
   regardless of the value of $\varphi$. The region with $|\Gamma|<1$
   becomes visible periodically when $\varphi<\varphic$, where
   $\cos\varphic \equiv - \Gamma$.}
 \label{fig:theta-gamma}
\end{figure}

The visible part of the surface location $(\theta,\varphi)$ relative
to the instantaneous observer's line-of-sight is illustrated in Figure
\ref{fig:theta-gamma} \citep[see also Figure 2 of][]{SSNB2022}. By
defining $\varphic(t)$ for $|\Gamma(t)| \leq 1$ as
\begin{eqnarray}
\label{eq:def-varphic}
\cos\varphic (t) \equiv - \Gamma(t) = -\cot\gamma(t)\cot\theta,
\end{eqnarray}
the instantaneously visible part of the surface is limited to
$-\varphic(t)<\varphi<\varphic(t)$.

Consider first the case of $0<\gamma<\pi/2$.
Then, equation (\ref{eq:Lt-0}) becomes
\begin{eqnarray}
\label{eq:Lt-0-2}
\hspace{-2cm}
\frac{\pi \overline{I}}{1-u_1-u_2} L_0(t)
&=& {\sin\gamma}
  \int_0^{\pi/2-\gamma}d\theta
  \int_{-\pi}^\pi d\varphi [\cos\varphi +\cot\gamma\cot\theta]\sin^2\theta
  \, {I(\theta)}\cr
&& \qquad \qquad +  {\sin\gamma}
  \int_{\pi/2-\gamma}^{\pi/2+\gamma}d\theta
  \int_{-\varphic}^{\varphic} d\varphi
  (\cos\varphi -\cos\varphic)\sin^2\theta   \, {I(\theta)}\cr
&& \hspace{-2cm} =2\pi {\cos\gamma} \int_0^{\pi/2-\gamma}
  \sin\theta\cos\theta {I(\theta)} \, d\theta + 2\sin\gamma
  \int_{\pi/2-\gamma}^{\pi/2+\gamma}
  (\sin\varphic -\varphic\cos\varphic)\sin^2\theta
  \, {I(\theta)} \, d\theta .
\end{eqnarray}
Using the following relations derived from
equation (\ref{eq:def-varphic}):
\begin{eqnarray}
\label{eq:theta-varphic}
\sin^2\theta = \frac{\cot^2\gamma}{\cos^2\varphic +\cot^2\gamma}, 
\qquad
d\theta = - \frac{\sin\varphic\cot\gamma}{\cos^2\varphic +\cot^2\gamma}
d\varphic,
\end{eqnarray}
the second term in equation (\ref{eq:Lt-0-2}) becomes an integral
in terms of $\varphic$ alone:
\begin{eqnarray}
  \int_{\pi/2-\gamma}^{\pi/2+\gamma}
  \sin^2\theta (\sin\varphic-\varphic\cos\varphic)\, d\theta
  =   \int_{\pi}^{0}
\cot^3\gamma \frac{\sin\varphic(\varphic\cos\varphic-\sin\varphic)}
  {(\cos^2\varphic +\cot^2\gamma)^2} \, d\varphic.
\end{eqnarray}
Finally, equation (\ref{eq:Lt-0-2}) reduces to
\begin{eqnarray}
\label{eq:Lt-0-2b}
\frac{\overline{I}}{1-u_1-u_2} L_0(t)
&&=2{\cos\gamma} \int_0^{\pi/2-\gamma}
\sin\theta\cos\theta {I(\theta)} \, d\theta \cr
&& \hspace{1cm} - \frac{2\sin\gamma \cot^3\gamma}{\pi}
  \int_{0}^{\pi} F_0(x;\gamma) I(\theta(x)) dx,
\end{eqnarray}
where 
\begin{eqnarray}
\label{eq:def-F0}
  F_0(x;\gamma) \equiv
  \frac{\sin x(x\cos x-\sin x)}{(\cos^2 x +\cot^2\gamma)^2} ,
  \qquad
  \cos[\theta(x)] \equiv \frac{\cos x}{\cos^2x +\cot^2\gamma}.
\end{eqnarray}

Similarly, equation (\ref{eq:Lt-1}) is rewritten as
\begin{eqnarray}
\label{eq:Lt-1-2}
\hspace{-2cm}
\frac{\overline{I}}{u_1+2u_2} L_1(t)
&=& \frac{\sin^2\gamma}{\pi}
  \int_0^{\pi/2-\gamma}d\theta
  \int_{-\pi}^\pi d\varphi [\cos\varphi +\cot\gamma\cot\theta]^2\sin^3\theta
  \, {I(\theta)}\cr
&& \qquad \qquad +  \frac{\sin^2\gamma}{\pi}
  \int_{\pi/2-\gamma}^{\pi/2+\gamma}d\theta
  \int_{-\varphic}^{\varphic} d\varphi
  (\cos\varphi -\cos\varphic)^2\sin^3\theta   \, {I(\theta)}\cr
 &=& \sin^2\gamma
  \int_0^{\pi/2-\gamma}
  \sin^3\theta {I(\theta)} \, d\theta + 
2 {\cos^2\gamma} \int_0^{\pi/2-\gamma}
  \sin\theta\cos^2\theta {I(\theta)} \, d\theta \cr
  && \qquad + \frac{\sin^2\gamma}{2\pi}
  \int_{\pi/2-\gamma}^{\pi/2+\gamma}
  (2\varphic\cos2\varphic - 3\sin2\varphic)\sin^3\theta
  \, {I(\theta)} \, d\theta \cr
 &=& \sin^2\gamma
  \int_0^{\pi/2-\gamma}  \sin^3\theta {I(\theta)} \, d\theta
  + 2{\cos^2\gamma} \int_0^{\pi/2-\gamma}
  \sin\theta\cos^2\theta {I(\theta)} \, d\theta \cr
  && \qquad - \frac{\sin^2\gamma\cot^4\gamma}{2\pi}
  \int_{0}^{\pi} F_1(x;\gamma) I(\theta(x)) dx ,
  \end{eqnarray}
where 
\begin{eqnarray}
\label{eq:def-F1}
  F_1(x;\gamma) \equiv
  \frac{\sin x(2x\cos 2x- 3\sin 2x +4x)}{(\cos^2 x +\cot^2\gamma)^{5/2}} .
\end{eqnarray}

Also equation (\ref{eq:Lt-2}) becomes
\begin{eqnarray}
\label{eq:Lt-2-2}
\hspace{-2cm}
-\frac{\overline{I}}{u_2} L_2(t)
&=& \frac{\sin^3\gamma}{\pi}
  \int_0^{\pi/2-\gamma}d\theta
  \int_{-\pi}^\pi d\varphi [\cos\varphi +\cot\gamma\cot\theta]^3\sin^4\theta
  \, {I(\theta)}\cr
&& \qquad +  \frac{\sin^3\gamma}{\pi}
  \int_{\pi/2-\gamma}^{\pi/2+\gamma}d\theta
  \int_{-\varphic}^{\varphic} d\varphi
  (\cos\varphi -\cos\varphic)^3\sin^4\theta   \, {I(\theta)}\cr
  &=& 3\sin^2\gamma\cos\gamma
  \int_0^{\pi/2-\gamma}
  \sin^3\theta \cos\theta {I(\theta)} \, d\theta + 
2 {\cos^3\gamma} \int_0^{\pi/2-\gamma}
  \sin\theta\cos^3\theta {I(\theta)} \, d\theta \cr
  && + \frac{\sin^3\gamma}{12\pi}
  \int_{\pi/2-\gamma}^{\pi/2+\gamma}
  [27\sin\varphic + 11\sin 3\varphic
  -6\varphic(9\cos\varphic +\cos3\varphic)]\sin^4\theta
  \, {I(\theta)} \, d\theta \cr
  &=& 3\sin^2\gamma\cos\gamma
  \int_0^{\pi/2-\gamma}
  \sin^3\theta \cos\theta {I(\theta)} \, d\theta + 
2{\cos^3\gamma} \int_0^{\pi/2-\gamma}
  \sin\theta\cos^3\theta {I(\theta)} \, d\theta \cr
  && \hspace{3.5cm} + \frac{\sin^3\gamma\cot^5\gamma}{12\pi}
  \int_{0}^{\pi} F_2(x;\gamma) I(\theta(x)) dx ,
\end{eqnarray}
where 
\begin{eqnarray}
\label{eq:def-F2}
  F_2(x;\gamma) \equiv
  \frac{\sin x[6x(9\cos x + \cos 3x) - 27 \sin x - 11\sin 3x ]}
       {(\cos^2 x +\cot^2\gamma)^{3}} .
\end{eqnarray}

In the same manner, we rewrite equations (\ref{eq:Lt-0}) to
(\ref{eq:Lt-2}) for $\pi/2<\gamma<\pi$ as follows.
\begin{eqnarray}
\label{eq:Lt-0-3}
\hspace{-1cm}
\frac{\overline{I}}{1-u_1-u_2} L_0(t)
&&=2 {\cos\gamma} \int_{3\pi/2-\gamma}^{\pi}
\sin\theta\cos\theta {I(\theta)} \, d\theta \cr
&& \hspace{1cm} - \frac{2\sin\gamma |\cot^3\gamma|}{\pi}
  \int_{0}^{\pi} F_0(x;\gamma) I(\theta(x)) dx,
\end{eqnarray}
\begin{eqnarray}
\label{eq:Lt-1-3}
\hspace{-1cm}
\frac{\overline{I}}{u_1+2u_2} L_1(t)
 &=&  \sin^2\gamma
  \int_{3\pi/2-\gamma}^{\pi} \sin^3\theta {I(\theta)} \, d\theta
  + 2 {\cos^2\gamma} \int_{3\pi/2-\gamma}^{\pi}
  \sin\theta\cos^2\theta {I(\theta)} \, d\theta \cr
  &&  \hspace{1cm} - \frac{\sin^2\gamma\cot^4\gamma}{2\pi}
  \int_{0}^{\pi} F_1(x;\gamma) I(\theta(x)) dx ,
\end{eqnarray}
\begin{eqnarray}
\label{eq:Lt-2-3}
\hspace{-1cm}
-\frac{\overline{I}}{u_2} L_2(t)
  &=& 3\sin^2\gamma\cos\gamma
  \int_{3\pi/2-\gamma}^{\pi}
  \sin^3\theta \cos\theta {I(\theta)} \, d\theta + 
2{\cos^3\gamma} \int_{3\pi/2-\gamma}^{\pi}
  \sin\theta\cos^3\theta {I(\theta)} \, d\theta \cr
  &&  \hspace{1.5cm} + \frac{\sin^3\gamma\cot^5\gamma}{12\pi}
  \int_{0}^{\pi} F_2(x;\gamma) I(\theta(x)) dx ,
\end{eqnarray}

\section{Application to simple models of the surface inhomogeneity
  \label{sec:application}}

\subsection{Cap-model: a constant-intensity circular spot
  \label{subsec:cap-model}}

\begin{figure}
 \begin{center}
 \includegraphics[width=14cm]{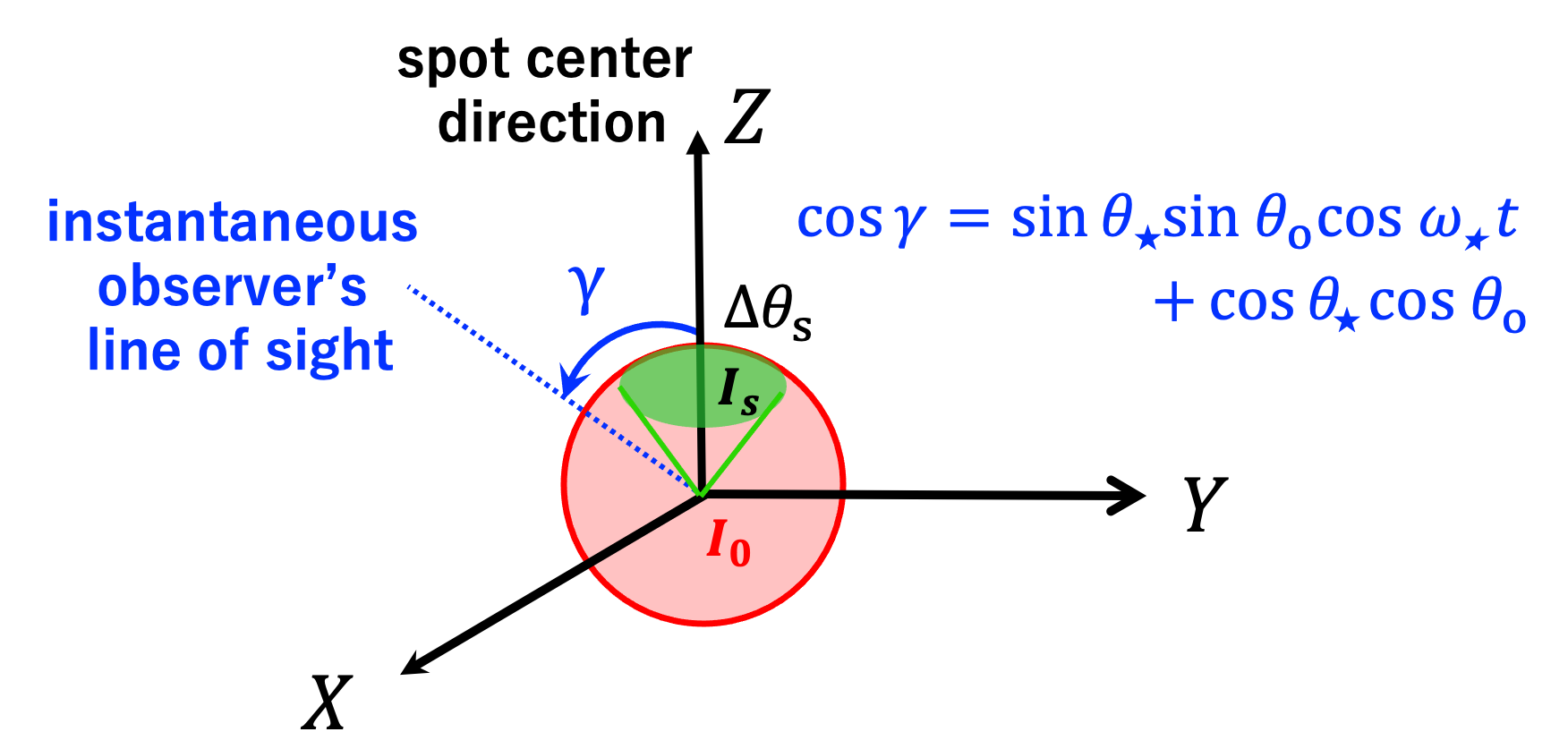} 
 \end{center}
 \caption{An instantaneous view of the spot configuration for the
   cap-model.  The spot is located at the pole region ($\theta<\dts$)
   with the constant intensity $I_{\rm s}$, while the intensity of the
   other surface is $I_0$.} \label{fig:spot-config}
\end{figure}

Let us consider a simple model (Figure \ref{fig:spot-config}) in which
the surface intensity is given as
\begin{eqnarray}
\label{eq:twoI-model}
  I(\theta) =
  \left\{
  \begin{array}{ll}
    I_s & ( 0<\theta<\dts )\\
    I_0 & ( \dts<\theta <\pi )
  \end{array}
  \right.
  .
\end{eqnarray}
This corresponds to a large spot of constant intensity $I_s$ around
the north pole on the otherwise homogeneous surface of intensity
$I_0$.  Without loss of generality, we assume that the size of the
spot is less than half of the entire surface ($\dts < \pi/2$); the
result for $\dts > \pi/2$ is obtained simply by exchanging $I_s$ and
$I_0$ and replacing $\dts$ with $\pi-\dts$ in the following
expressions.

We compute equation (\ref{eq:Lt-0-2b}) without limb darkening (the
derivation is given in appendix \ref{app:derivation}), and summarize
the final expressions below.
\begin{description}
\item[(a)] $0<\gamma<\pi/2$ and $\dts<\pi/2-\gamma$:
\begin{eqnarray}
\label{eq:L0a}
L_{\rm 0a}(t) = \frac{I_0}{\overline{I}}
- \frac{I_0-I_s}{\overline{I}}\sin^2\dts \cos\gamma .
\end{eqnarray}
\item[(b)] $0<\gamma<\pi/2$ and $\pi/2-\gamma<\dts<\pi/2$:
\begin{eqnarray}
\label{eq:L0b}
L_{\rm 0b}(t)
= \frac{I_0}{\overline{I}} 
- 2\sin\gamma\cot^3\gamma \frac{(I_0-I_s)}{\pi\overline{I}}
\int_{\varphics}^{\pi}F_0(x;\gamma) dx ,
\end{eqnarray}
where  $\cos\varphics \equiv -\cot\gamma\cot\dts$.
\item[(c)] $\pi/2<\gamma<\pi$ and $\dts<\gamma-\pi/2$:
\begin{eqnarray}
\label{eq:L0c}
L_{\rm 0c}(t) =  \frac{I_0}{\overline{I}} .
\end{eqnarray}
\item[(d)] $\pi/2<\gamma<\pi$ and $\gamma-\pi/2 <\dts<\pi/2$:
\begin{eqnarray}
\label{eq:L0d}
L_{\rm 0d}(t)
=  \frac{I_0}{\overline{I}} 
- 2\sin\gamma\cot^3\gamma \frac{(I_0-I_s)}{\pi\overline{I}}
\int_0^{\varphics}F_0(x;\gamma) dx .
\end{eqnarray}
\end{description}

The case (a) corresponds to the situation that the whole spot is
visible to the observer, and thus the modulation is proportional to
the projected area of the spot ($\propto\cos\gamma$).  This clearly
indicates that the resulting modulation is proportional to
$\cos\omegas t$; see equation (\ref{eq:gamma}). In contrast, the spot
is completely invisible and there is no observed modulation in the
case (c). In the other two cases (b) and (d), the part of the spot is
periodically visible, and the corresponding lightcurves become a
sinusoidal form truncated at upper/lower amplitudes.

We note that the cap-model can be generalized to a star with multiple
large spots in a straightforward manner by summing up the photometric
variations due to the $i$-th spot with $I_{{\rm s}, i}$,
$\theta_{{\rm s},i}(t)$, $\Delta\theta_{{\rm s},i}$, and the initial
phase of $\varphi_{i}$.

\subsection{p-model: an inhomogeneous surface with a globally varying
  intensity
  \label{subsec:p-model}}

Another simple model that we consider here assumes the parameterized
surface intensity $I(\theta)$:
\begin{eqnarray}
\label{eq:p-model}
  I(\theta) = I_0 - \dI \cos^p\theta  ,
\end{eqnarray}
where $I_0$ is the intensity at the equator ($\theta=\pi/2$), and
$\dI$ is the amplitude of the intensity variation on the surface.

If $p$ is an odd integer, $I(\theta)$ varies monotonically from
$I_0-\dI$ (at $\theta=0$) to $I_0+\dI$ (at $\theta=\pi$), which is
roughly consistent with previous data
\citep[e.g.,][]{Landolfi1993,Valyavin2008,Valyavin2011, Valyavin2014}.
In this case, equation (\ref{eq:Lt-0-2b}) is rewritten as
\begin{eqnarray}
  \label{eq:Lt-p-odd}
  L_{\rm 0p}(t)
&& =
 \frac{I_0}{\overline{I}} 
  - \frac{\dI}{\overline{I}}
 \frac{2(1-\sin^{p+2}\gamma)\cos\gamma}{p+2} \cr
 && \qquad\qquad
 - \frac{\dI}{\overline{I}}\frac{2\sin\gamma\cot^3\gamma}{\pi}
 \int_{0}^{\pi} F_0(x;\gamma)
 \left(\frac{\cos x}{\sqrt{\cos^2 x+ \cot^2\gamma}}\right)^p dx ,
\end{eqnarray}
where the second line is obtained from equation (\ref{eq:int-F0}).
Interestingly, after integrating by parts, we find the analytical
result:
\begin{eqnarray}
 \int_{0}^{\pi} F_0(x;\gamma)
 \left(\frac{\cos x}{\sqrt{\cos^2 x+ \cot^2\gamma}}\right)^p dx
 = \frac{\pi}{p+2} \frac{\sin^{p+2}\gamma}{\cot^2\gamma} ,
\end{eqnarray}
if $p$ is an odd integer. Thus, equation (\ref{eq:Lt-p-odd}) reduces
to the following simple expression:
\begin{eqnarray}
  \label{eq:Lt-p-odd-final}
  L_{\rm 0p}(t)
 = \frac{I_0}{\overline{I}} 
  - \frac{\dI}{\overline{I}}
 \frac{2}{p+2} \cos\gamma.
\end{eqnarray}
As in the cap-model, it has a modulation component proportional
to $\cos\gamma$, and thus to $\cos\omegas t$, but the parameter $p$
cannot be measured because it is completely degenerate with the
modulation amplitude $\dI$, as long as it is an odd integer.

\subsection{Lightcurve variations without limb darkening
\label{subsec:L0}}

\begin{figure}
  \begin{center}
\hspace*{-1cm}\includegraphics[width=13cm]{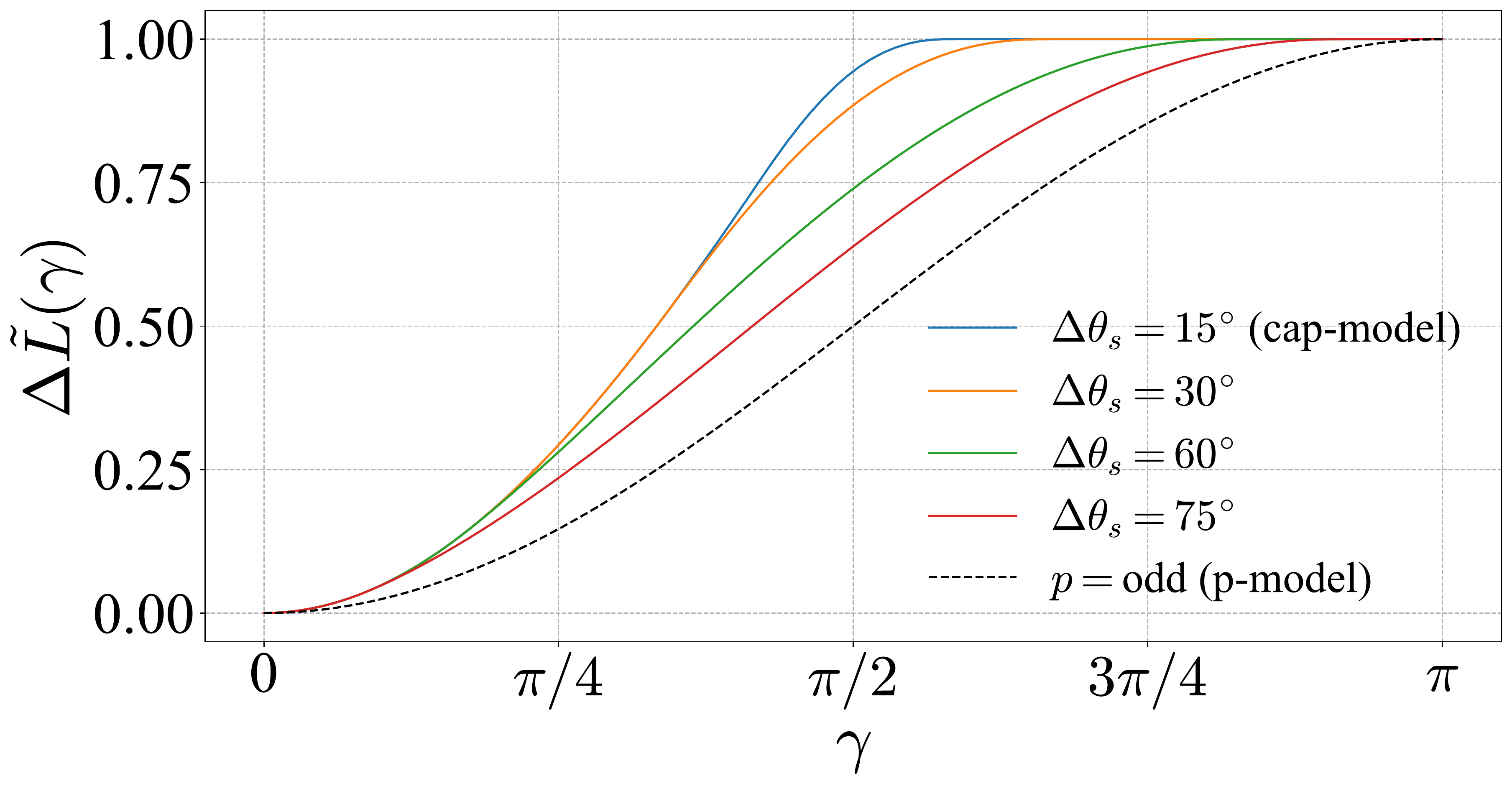}
\hspace*{0.5cm}\includegraphics[width=12cm]{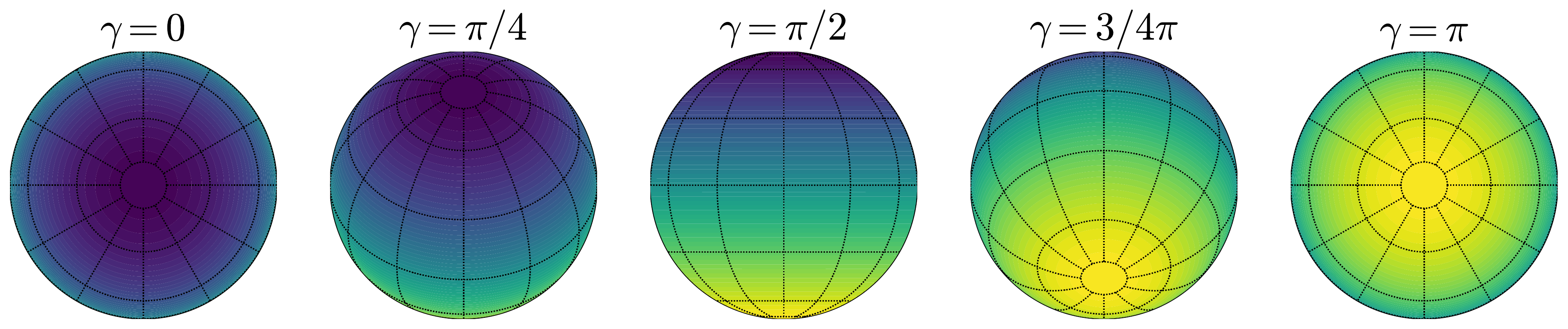} 
 \end{center}
  \caption{Scaled lightcurve variation $\Delta \tilde{L}(\gamma) $
    against $\gamma$ for $I_{\rm s}<I_0$. The four solid lines
    correspond to the cap-model with $\dts=15^\circ$, $30^\circ$,
    $60^\circ$, and $75^\circ$, while the dashed line corresponds to
    the p-model. The bottom panel shows a schematic illustration of
    the stellar surface at different phases for the p-model.
     \label{fig:L-gamma}}
\end{figure}

In our modeling, the time-dependence of the lightcurve $L(t)$ is
completely specified by that of $\gamma(t)$. Thus we define the scaled
lightcurve variation in terms of $\gamma$:
\begin{eqnarray}
\label{eq:scaled-dLgamma}
 \Delta \tilde{L}(\gamma)
\equiv \frac{L(\gamma) - L(\gamma=0)}{L(\gamma=\pi)-L(\gamma=0)}, 
\end{eqnarray}
which is plotted in Figure \ref{fig:L-gamma} for our cap-model and
p-model without limb darkening.  For the p-model, $\Delta
\tilde{L}(\gamma)$ simply reduces to $(1-\cos\gamma)/2$ regardless of
the value of $p$, as shown in equation (\ref{eq:Lt-p-odd-final}). In
contrast, $\Delta \tilde{L}(\gamma)$ depends on the value of $\dts$ as
well; see equations (\ref{eq:L0a}) to (\ref{eq:L0d}).

\begin{figure}
  \begin{center}
    \includegraphics[width=12cm]{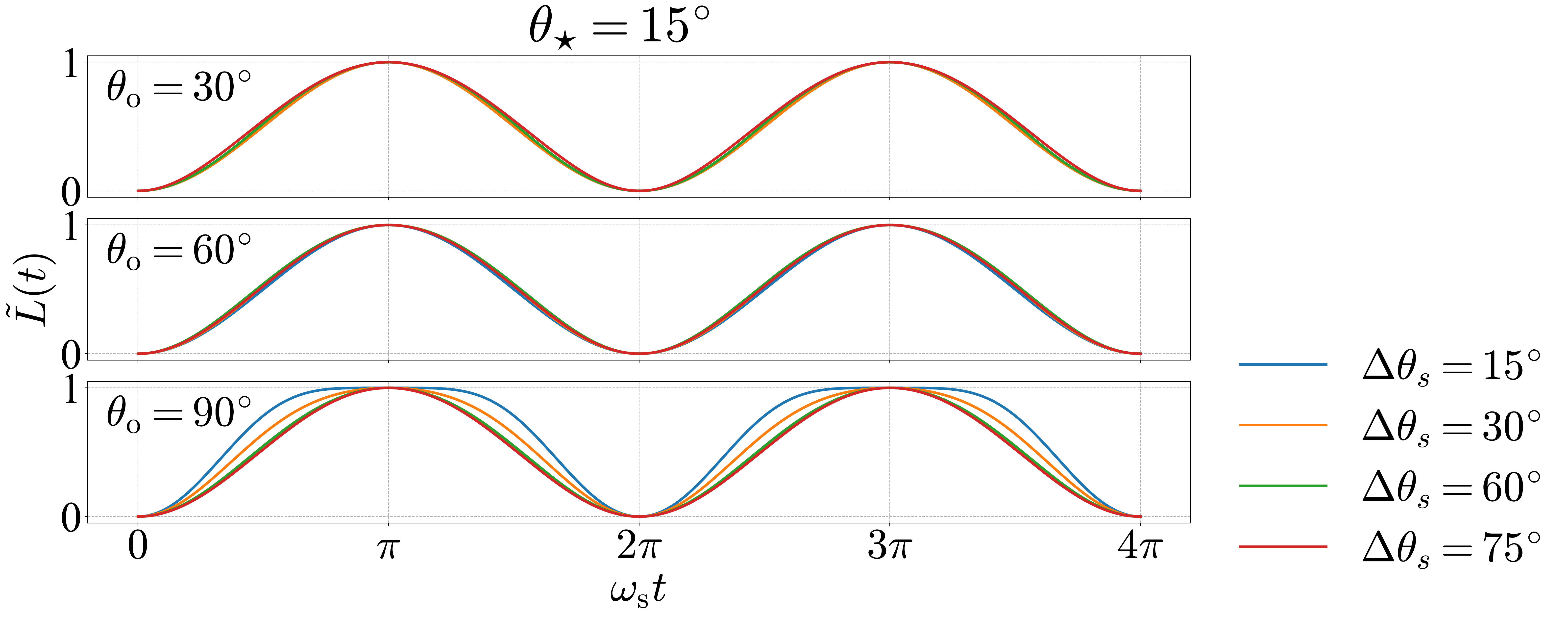} 
    \includegraphics[width=12cm]{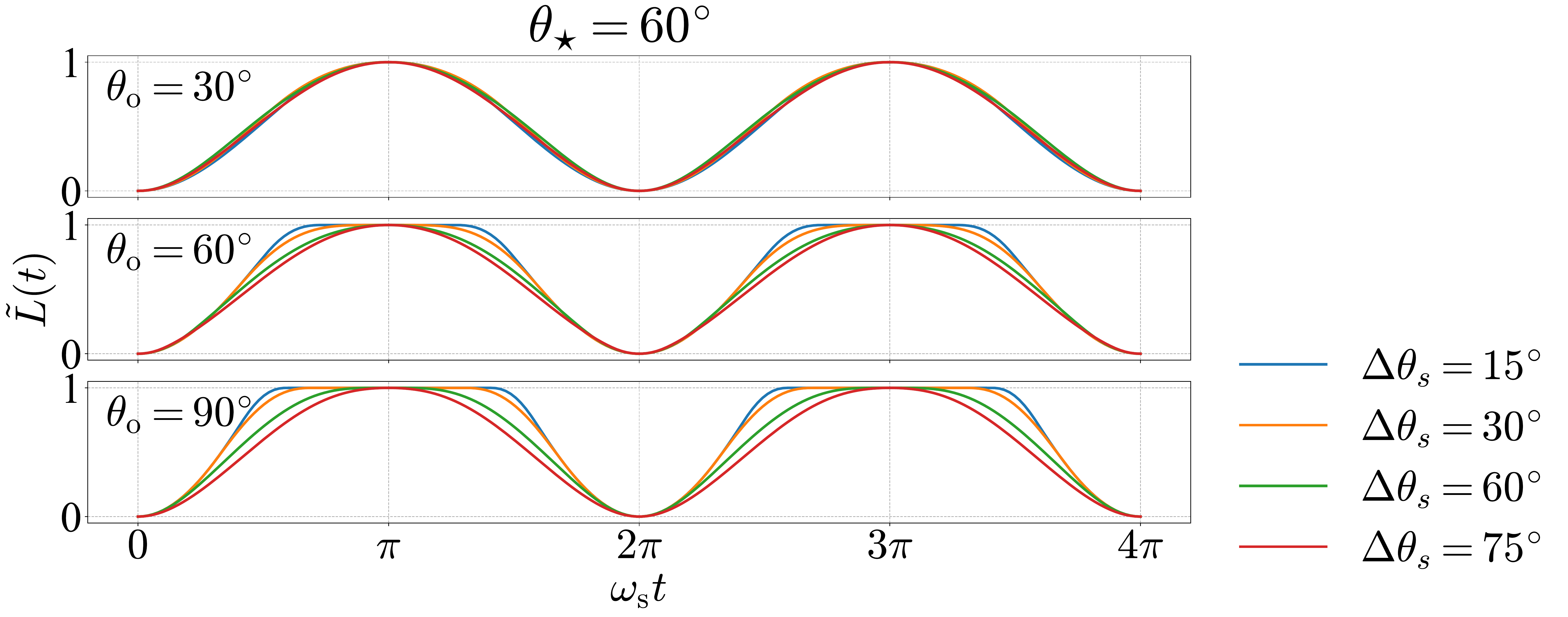}
    \includegraphics[width=12cm]{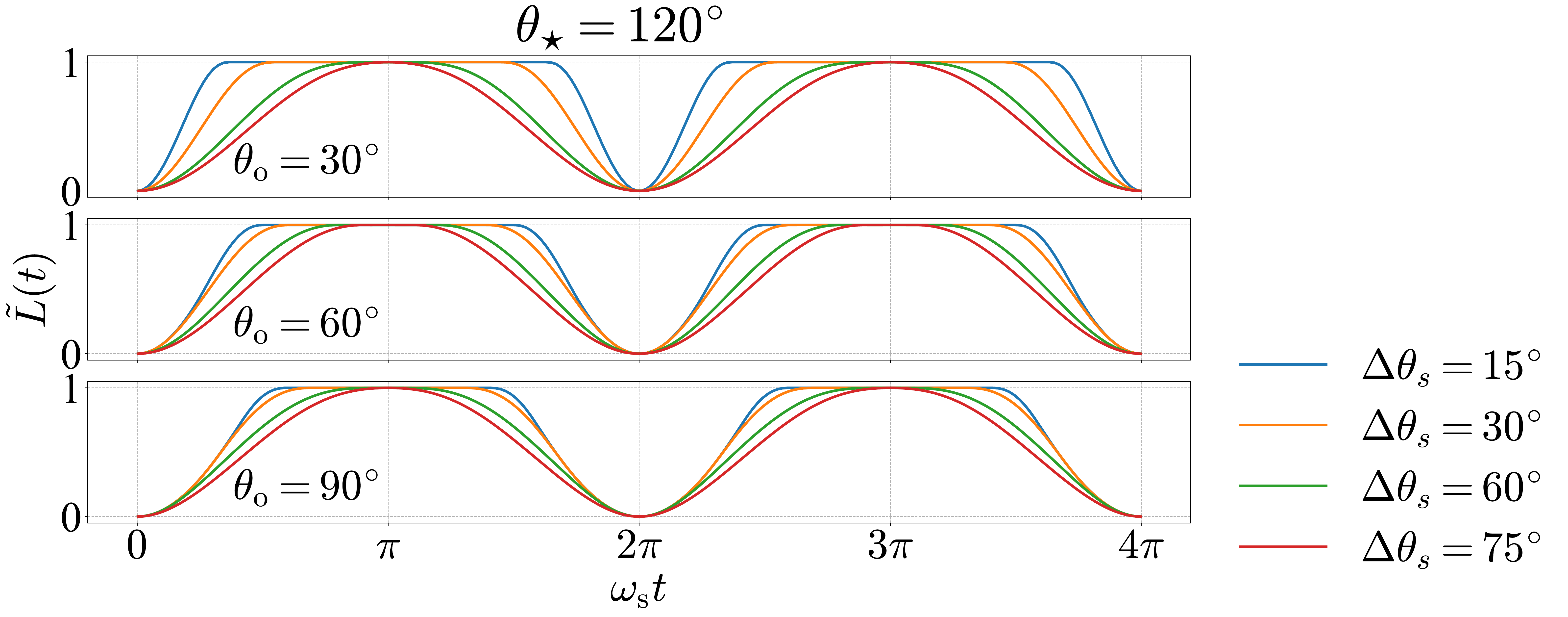}
 \end{center}
  \caption{Scaled lightcurve variation $\tilde{L}(t)$ for the
    cap-model against the orbital phase ($\omegas t$) with
      $I_{\rm s}<I_0$.  The top, middle, and bottom panels correspond
    to $\thetaS=15^\circ$, $60^\circ$, and $120^\circ$ respectively,
    for $\thetao= 30^\circ$, $60^\circ$, and $90^\circ$. Each panel
    has four curves corresponding to the spot size of $\dts=15^\circ$,
    $30^\circ$, $60^\circ$, and $75^\circ$.
    \label{fig:Lt-cap}}
\end{figure}

In reality, however, the value of $\gamma$ is not known to the
observer, and $\Delta \tilde{L}(\gamma)$ is not directly
observable. Instead, we define the scaled lightcurve variation in
terms of $t$:
\begin{eqnarray}
\label{eq:scaled-dLt}
 \tilde{L}(t) \equiv \frac{L(\omegas t) - L_{\rm 0a}(\omegas t=0)}
        {L_{\rm 0a}(\omegas t=\pi)-L_{\rm 0a}(\omegas t=0)},
\end{eqnarray}
which can be evaluated by fitting equation (\ref{eq:L0a}) to the
sinusoidal portion of the measured data.  Figure \ref{fig:Lt-cap}
plots $\tilde{L}(t)$ for the cap-model against $\omegas t$ without
limb darkening.

There is a certain degeneracy among parameters in the cap-model. If
the lightcurve becomes saturated corresponding to equation
(\ref{eq:L0c}), we can measure the threshold value of $\omegas
t=\theta_{\rm crit}$ which gives
\begin{eqnarray}
\label{eq:dts-thetacrit}
\sin\dts = |\cos\gamma_{\rm crit}| \equiv |\sin\thetaS \sin\thetao
\cos\theta_{\rm crit} + \cos\thetaS \cos\thetao | .
\end{eqnarray}
Then fitting to equation (\ref{eq:L0a}) with equation
(\ref{eq:dts-thetacrit}) provides a relation among $I_s/I_0$,
$\thetaS$ and $\thetao$:
\begin{eqnarray}
\label{eq:L0a-fit}
L_{\rm 0a}(t) &&= \frac{I_0}{\overline{I}}
 - \frac{I_0-I_s}{\overline{I}}
  \sin^2\dts \left(\sin\thetaS \sin\thetao \cos\omegas t
  + \cos\thetaS \cos\thetao\right) \cr
  &&= 
  \left(\frac{I_0}{\overline{I}}
  -\frac{I_0-I_s}{\overline{I}}\sin^2\dts\cos\thetaS \cos\thetao  \right)
  \cr
  && \qquad \times
  \left[1 -\frac{(1-I_s/I_0)\sin^2\dts \sin\thetaS \sin\thetao}
       {1-(1-I_s/I_0)\sin^2\dts\cos\thetaS \cos\thetao}
        \cos\omegas t \right].
\end{eqnarray}
The above equation explicitly indicates the degeneracy of the
parameters involved in the amplitude of the sinusoidal curve.
Therefore, the departure from the purely sinusoidal curve (see Figure
\ref{fig:Lt-cap}) corresponding to equations (\ref{eq:L0b}) to
(\ref{eq:L0d}) is important for the accurate determinations of the
model parameters. We will discuss the parameter degeneracy in
\S\ref{subsec:application2ZTF} below.

\subsection{Limb darkening effect for the p-model \label{subsec:LD-p}}

We compute the effect of limb darkening for the p-model.  Substituting
equation (\ref{eq:p-model}) to equations (\ref{eq:Lt-1-2}),
(\ref{eq:Lt-2-2}), (\ref{eq:Lt-1-3}), and (\ref{eq:Lt-2-3}), we find
\begin{eqnarray}
\label{eq:Lt-1-pmodel}
L_{\rm 1p}(\gamma)
&=& (u_1+2u_2) \left[\frac{2}{3}\frac{I_0}{\overline{I}}
  - \frac{\dI}{\overline{I}} J_{\rm 1p}(\gamma)\right], \\
\label{eq:Lt-2-pmodel}
L_{\rm 2p}(\gamma)
&=& -u_2 \left[\frac{1}{2}\frac{I_0}{\overline{I}}
  -\frac{\dI}{\overline{I}} J_{\rm 2p}(\gamma)\right],
  \end{eqnarray}
where
\begin{eqnarray}
\label{eq:J_1p}
J_{\rm 1p}(\gamma) && = 
\frac{2+2p\cos^2\gamma-\sin^{p+3}\gamma[3(p+1)\cos^2\gamma+2]}
     {(p+1)(p+3)} \frac{\cos\gamma}{|\cos\gamma|}\cr
&& \qquad\qquad\qquad
- \frac{\sin^2\gamma\cot^5\gamma}{2\pi|\cot\gamma|}
\int_{0}^{\pi} F_1(x;\gamma)
\left(\frac{\cos x}{\sqrt{\cos^2 x+ \cot^2\gamma}}\right)^p dx
\\
\label{eq:J_2p}
J_{\rm 2p}(\gamma) && =
\frac{6+ 2(p-1)\cos^2\gamma -\sin^{p+4}\gamma [5(p+2)\cos^2\gamma+6]}
      {(p+2)(p+4)} \cos\gamma \cr
&& \qquad\qquad\qquad
+ \frac{\sin^3\gamma\cot^5\gamma}{12\pi}
\int_{0}^{\pi} F_2(x;\gamma)
\left(\frac{\cos x}{\sqrt{\cos^2 x+ \cot^2\gamma}}\right)^p dx \cr
&& = \frac{6+ 2(p-1)\cos^2\gamma}{(p+2)(p+4)} \cos\gamma .
  \end{eqnarray}
Thus, the normalized lightcurve for the p-model including
quadratic limb darkening is given by
\begin{eqnarray}
  \label{eq:Lt-p-odd-total}
   L_{\rm p}(\gamma)
   &&= \frac{I_0}{\overline{I}}
   \left[(1-u_1-u_2) + \frac{2(u_1+2u_2)}{3} -\frac{u_2}{2}\right]\cr
   && \qquad   - \frac{\dI}{\overline{I}}
   \left[ \frac{2(1-u_1-u_2)}{p+2} \cos\gamma 
    + (u_1+2u_2)J_{\rm 1p}(\gamma) -u_2 J_{\rm 2p}(\gamma) \right],
\end{eqnarray}
where $p$ is an odd integer. 

\begin{figure}
  \begin{center}
 \includegraphics[width=12cm]{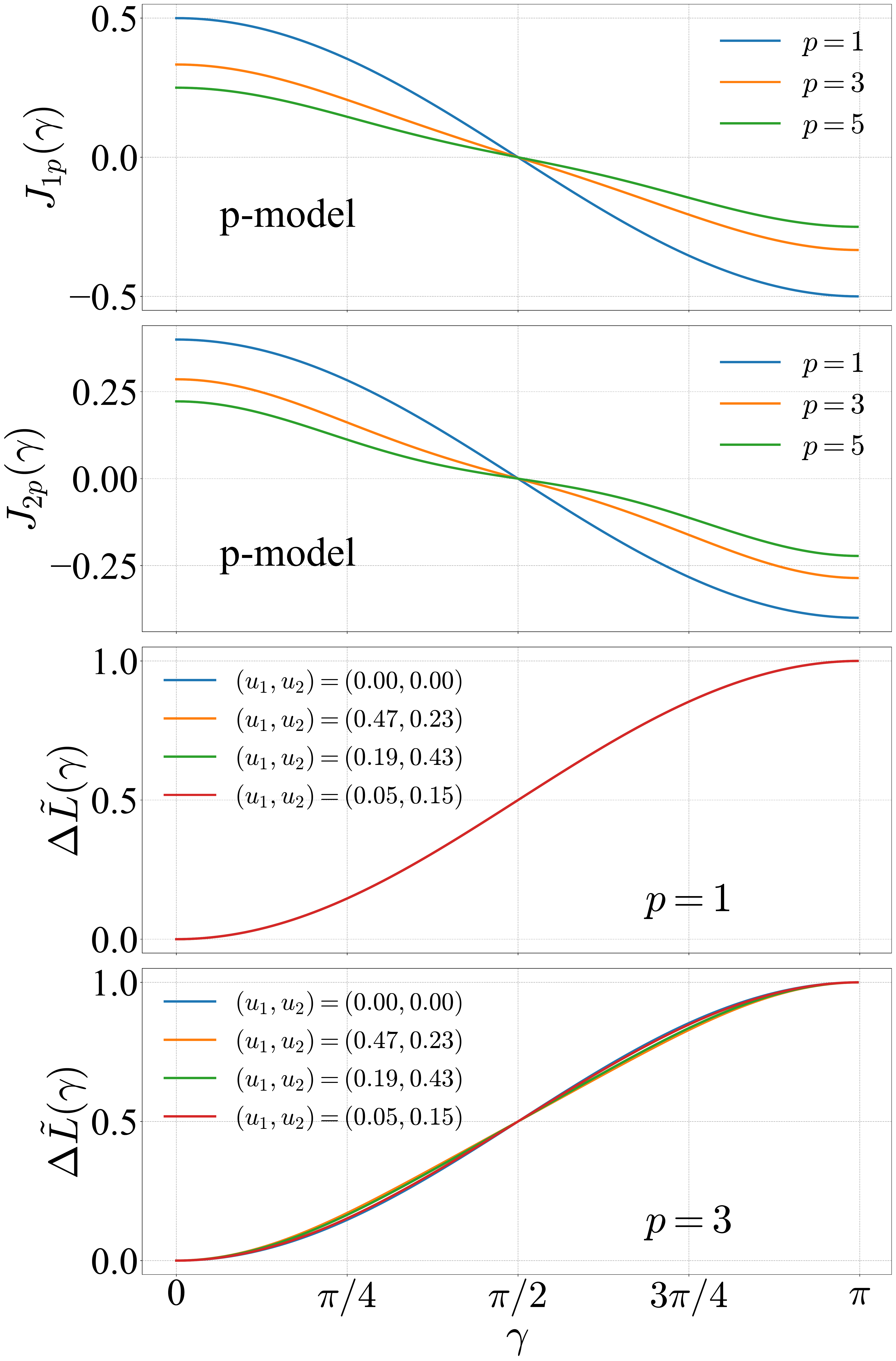} 
 \end{center}
  \caption{Effect of limb darkening for the p-model.  The upper two
    panels are $J_{\rm 1p}$ and $J_{\rm 2p}$ against $\gamma$ with
    different values of $p$. The lower two panels plot the
    corresponding scaled lightcurve variations $ \Delta \tilde{L}_{\rm
      p}$ for $p=1$ and 3, adopting the different sets of the limb
    darkening coefficients $(u_1,\; u_2)$.
        \label{fig:Lt-LD-pmodel}}
\end{figure}

The upper two panels in Figure \ref{fig:Lt-LD-pmodel} show $J_{\rm
  1p}$ and $J_{\rm 2p}$ against $\gamma$ for the p-model with $p=1$
(blue), 3 (orange) and 5 (green). The corresponding scaled lightcurve
variations $ \Delta \tilde{L}_{\rm p} (\gamma)$, computed from
equations (\ref{eq:scaled-dLgamma}) and (\ref{eq:Lt-p-odd-total}) are
plotted in the lower two panels for $p=1$ and 3, adopting the limb
darkening coefficients of $(u_1,\; u_2)= (0.47,\; 0.23)$ for the Sun,
and $(0.19,\;0.43)$ and $(0.05,\;0.15)$ for WDs with $(T_{\rm eff},\;
\log g)=(8 \times 10^{3} {\rm K}, \;9.2)$, and $(4.6 \times 10^{4}
     {\rm K},\; 9.5)$, respectively \citep{Kilic2021,Caiazzo2021}.
     For reference, $ \Delta \tilde{L}_{\rm p} (\gamma)$ without limb
     darkening is plotted as well, while it is barely distinguishable
     from the other curves.

In the case of $p=1$, the scaled lightcurves seem to be identical to
$(1-\cos \gamma)/2$, independently of the limb-darkening coefficients.
This implies that both $J_{\rm 1p}$ and $J_{\rm 2p}$ are proportional
to $\cos\gamma$ for $p=1$. Equation (\ref{eq:J_2p}) indeed shows that
it is the case for $J_{\rm 2p}$(p=1). We believe that it is also the
case for $J_{\rm 1p}(p=1)$, while we are not yet able to prove it
analytically.  The scaled lightcurves with $p=3$, on the other hand,
show very weak dependence on the limb-darkening coefficients, but they
are still very well approximated by $(1-\cos \gamma)/2$. Therefore, we
conclude that the limb-darkening effect is practically negligible in
the p-model, and expect that it is the case for the cap-model as well.

\subsection{Fitting the cap-model to the phase-folded lightcurves
  of a fast-spinning white dwarf ZTF J190132.9+145808.7 
  \label{subsec:application2ZTF}}

As an application of our current modeling, we attempt to fit the
lightcurve of a fast-spinning white dwarf ZTF J190132.9+145808.7. We
download the reduced photometric
lightcurves\footnote{https://github.com/ilac/ZTF-J1901-1458} that are
made publicly available by \citet{Caiazzo2021}.  The total duration of
the data is $\sim 80$ minutes, and the cadence is $\sim 1$ second. We
use the data in the blue band alone, and omit the last 200 data points
because they seem to be significantly contaminated by outliers. The
long-term trend is then removed using the Savitzky-Golay filter
\citep{SGfilter_1964}, implemented as {\tt LightCurve.flatten} in {\tt
  lightkurve} modules \citep{lightkurve2018}. We adopt a window width
of $1500$ seconds, and the data are folded according to the rotational
period of $416.097$ seconds. Finally, we average the data over the bin
width of $1$ second.

\begin{figure}
  \begin{center}
 \includegraphics[width=7cm]{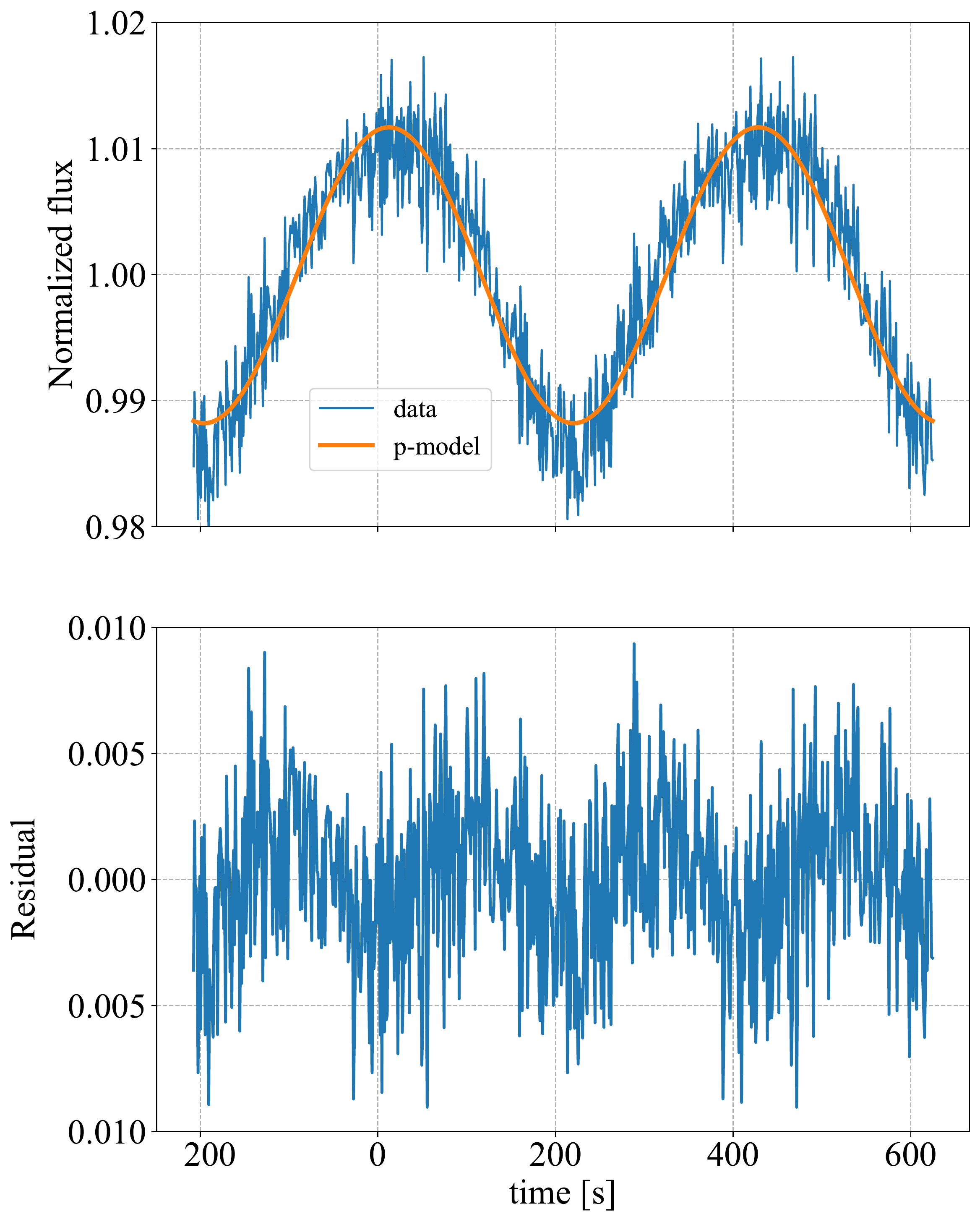} 
 \includegraphics[width=7cm]{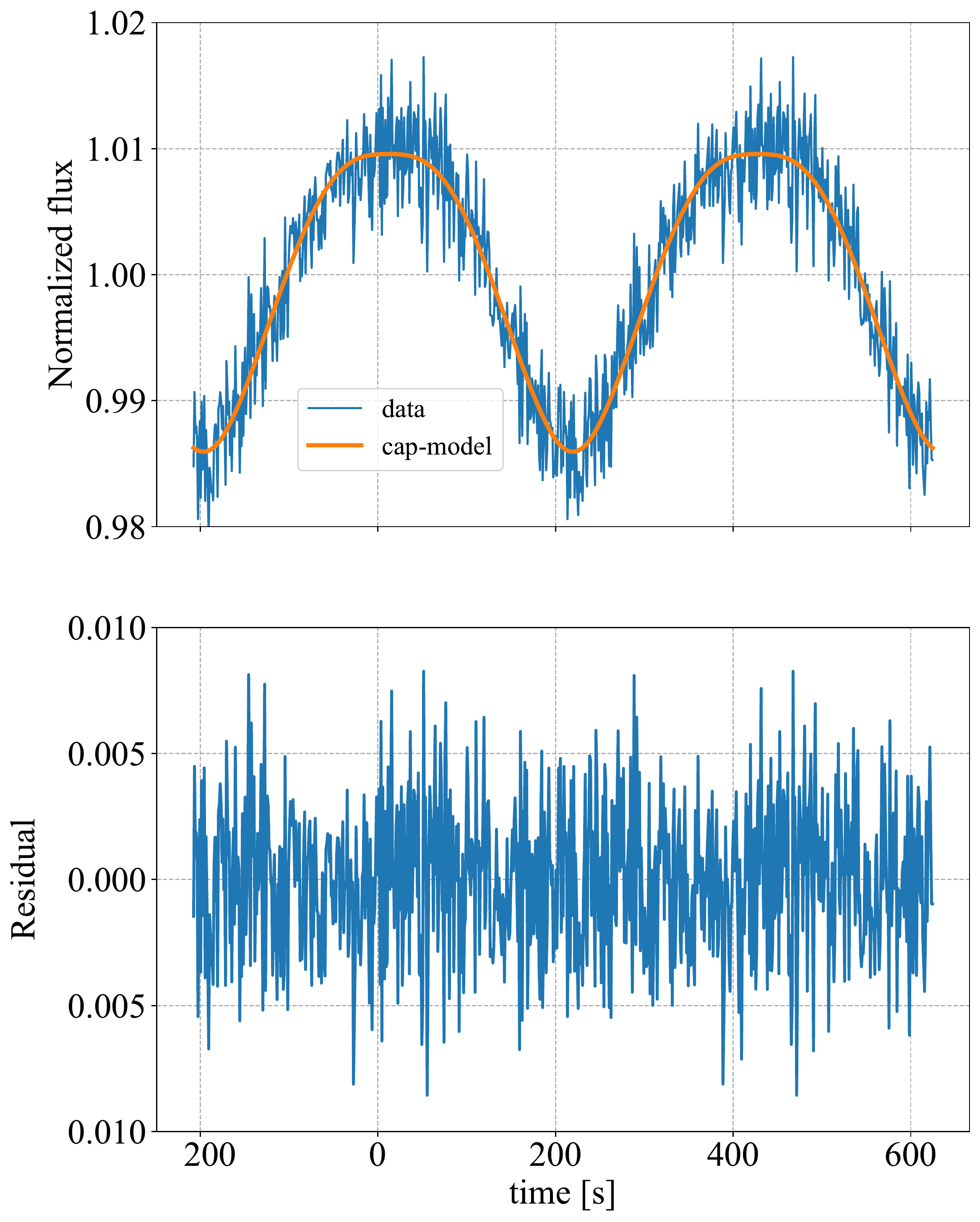} 
 \end{center}
  \caption{Best-fits to the phase-folded normalized lightcurve of ZTF
    J190132.9+145808.7, and their residuals for p-model (left) and
    cap-model (right).
    \label{fig:fit2ZTF}}
\end{figure}

As shown in the above subsection, the limb darkening effect is
practically negligible. Thus, we ignore it in the present
analysis. Figure \ref{fig:fit2ZTF} shows the best-fits and the
residuals for the p-model (left) and the cap-model (right) from the
nonlinear least square regression. While a purely sinusoidal curve
predicted from the p-model does not fit the data, the cap-model is
consistent with the feature of the observed lightcurve. Thus, we
proceed to perform a Markov Chain Monte Carlo (MCMC) ensemble sampler
for the cap-model using {\em EMCEE} \citep[][]{2013PASP..125..306F}.

\begin{figure}
  \begin{center}
 \includegraphics[width=17cm]{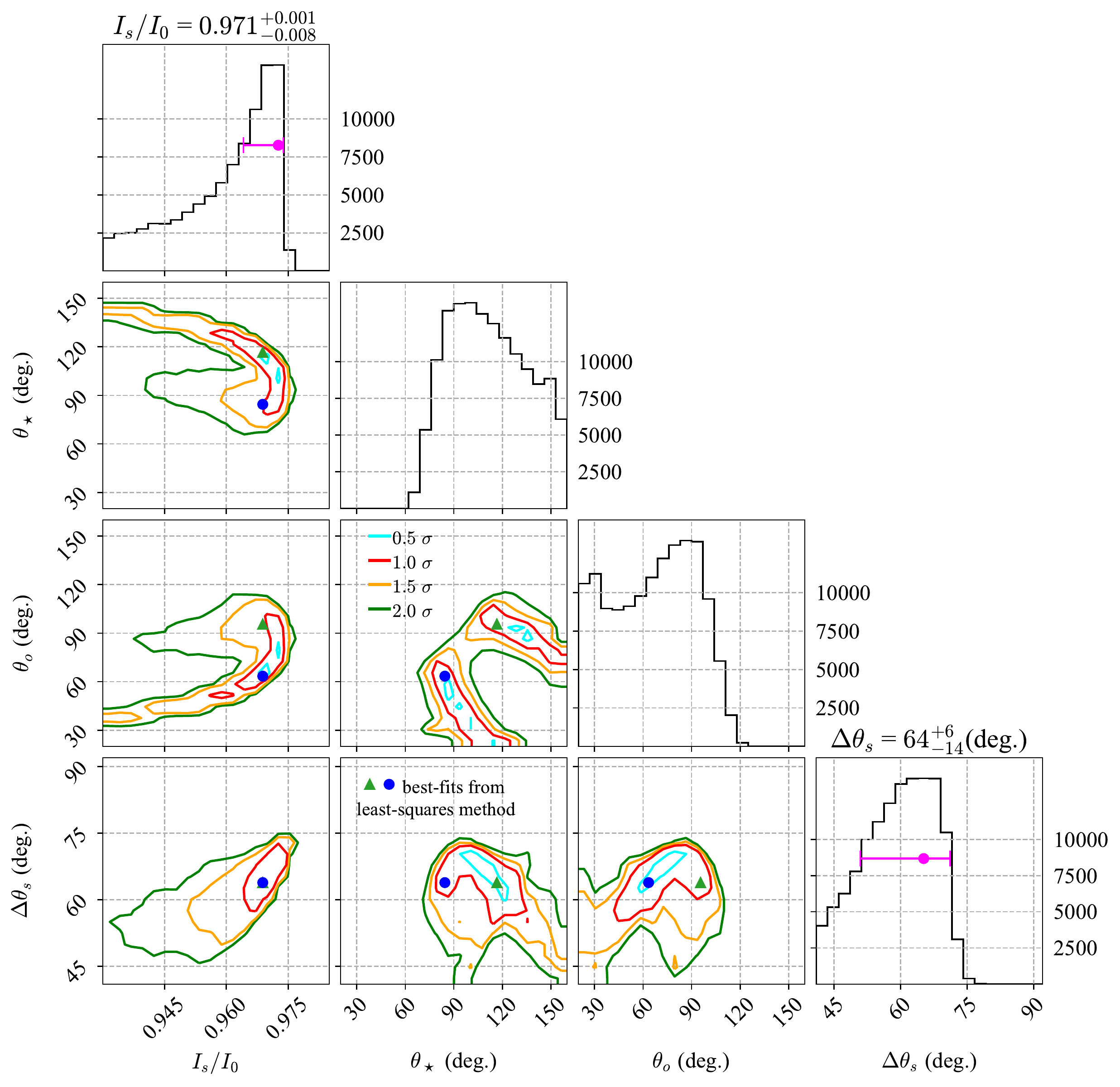} 
 \end{center}
  \caption{Posterior distribution among the four parameters,
    $I_s/I_o$, $\dts$, $\thetao$ and $\thetaS$, from the MCMC analysis
    in the cap-model. The observed lightcurve is averaged over a bin
    size of 5 seconds. The filled circle and triangle indicate
    solutions obtained from the nonlinear least square fit. Different
    curve correspond to the contours of $0.5 \sigma$ (cyan) , $1.0
    \sigma$ (red), $1.5\sigma$ (orange), and $2.0\sigma$ (green)
    levels for two-dimensional histograms, respectively.  The magenta
    symbols with error-bars indicate our best-fit parameters.  The
    figure is generated using the {\em CORNER} package
    \citep[][]{corner}.
      \label{fig:posterior}}
\end{figure}

\begin{figure}
  \begin{center}
 \includegraphics[width=12cm]{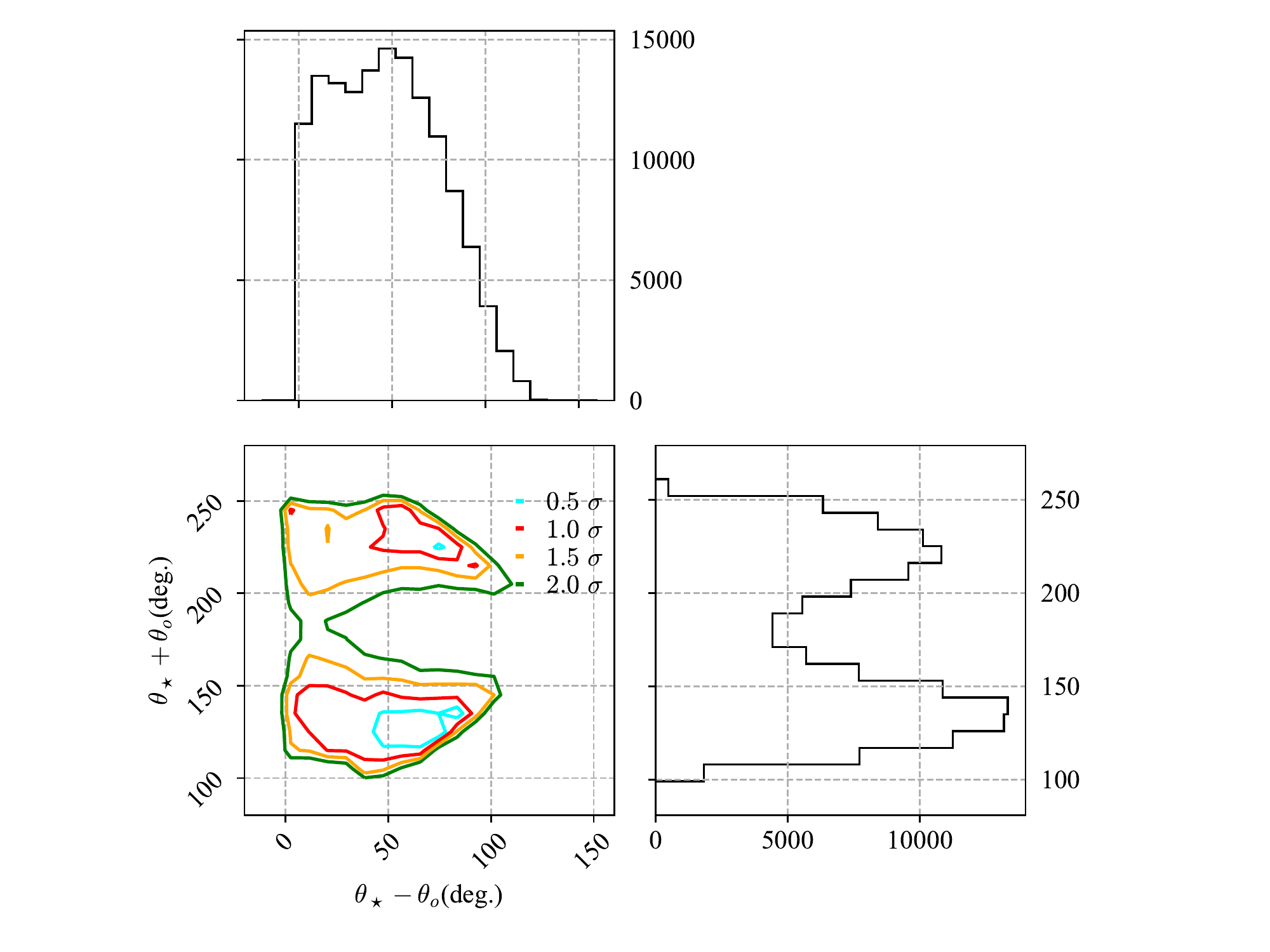} 
 \end{center}
  \caption{Same of Figure \ref{fig:posterior}, but the posterior
    distribution of $\thetaS - \thetao$ and $\thetaS + \thetao$ in
    the cap-model.
      \label{fig:posterior2}}
\end{figure}

  The likelihood function is taken to be proportional to
  $\mathrm{exp}(-\chi^2)$:
\begin{equation}
 \chi^2 = \sum_{i} \frac{(L_{\mathrm{cap},i} - L_{\mathrm {obs},i})^2}{\sigma_i^2},
\end{equation}
where $L_{\mathrm{cap},i}$ is the cap-model prediction,
$L_{\mathrm{obs},i}$ is the observed data, and $\sigma_i^2$ is the
variance of the data in the $i$-th bin.  The predicted lightcurve of
the cap-model is normalized to fit the amplitude of the observed light
curve.  We adopt uniform priors for the four free parameters over the
ranges of $0.8 < I_s / I_o < 1.2$, $0^\circ < \dts < 90^\circ$, and
$0^\circ < \thetao, ~\thetaS < 180^\circ$.

The posterior distributions for $I_s/I_o$, $\dts$, $\thetao$ and
  $\thetaS$ from the MCMC analysis are shown in Figure
  \ref{fig:posterior}, together with their marginalized probability
  density functions. The degree of the parameter degeneracy is clearly
  visible in Figure \ref{fig:posterior}. The amplitude of the
  photometric variation and the departure from the sinusoidal curve
  are specified by $I_s/I_o$ and $\dts$, respectively. Thus, these two
  parameters are well constrained from the fit.  We define the
best-fit parameters of $I_s/I_o$ and $\dts$ by the mode and the
corresponding $1\sigma$ errors from the histograms in Figure
\ref{fig:posterior}.

In contrast, the remaining two parameters, $\thetaS$ and
  $\thetao$, are intrinsically degenerate, which can be understood
  from the fact that $\gamma$ is invariant by exchanging $\thetao$ and
  $\thetaS$. Therefore, we put a constraint of $\thetaS > \thetao$ in
  Figure \ref{fig:posterior}.  Even under the constraint, there are
  two peaks, roughly corresponding to $(\thetao, \thetaS)$ $\approx
  (60^\circ, 90^\circ)$ and $(90^\circ, 120^\circ)$. In reality,
  however, these two are the same solution because $\thetaS= 60^\circ$
  and $120^\circ$ lead to the identical lightcurve if
  $\thetao=90^\circ$ as shown in Figure \ref{fig:Lt-cap}. Thus, the
  peak at $(\thetao, \thetaS)$ $\approx (90^\circ, 120^\circ)$ is the
  same at $(90^\circ, 60^\circ)$, which is simply equivalent to the
  other peak at $(\thetao, \thetaS)$ $\approx (60^\circ,
  90^\circ)$.

This parameter degeneracy is shown more clearly in Figure
\ref{fig:posterior2}, where we plot the posterior distribution on
the $\thetaS-\thetao$ vs $\thetaS+\thetao$ plane.  Therefore, the
best-fits values of $\thetao$ and $\thetaS$ are not easy to specify,
and we decide to adopt their modes and the associated $1\sigma$ errors
from the two-dimensional contours on the $\thetao$ -- $\thetaS$ plane
in Figure \ref{fig:posterior}. Having such degeneracy in mind, we
summarize the best-fit parameters of the cap-model for ZTF
J190132.9+145808.7 as follows:
\begin{eqnarray}
\label{eq:ZTF-param}
I_s/I_0 && = 0.971^{+0.001}_{-0.008}, \qquad
\dts = 64^{+6}_{-14}~\mathrm{deg.},\cr
(\thetao, \thetaS) && = (55^{+11}_{-31}~\mathrm{deg.}, 85^{+31}_{-7}~\mathrm{deg.}) 
~{\rm or} ~(90^{+18}_{-12}~\mathrm{deg.}, 120^{+24}_{-14}~\mathrm{deg.}).
\end{eqnarray}
The above solutions are indeed quite interesting. If $\thetao\approx
60^\circ$, the spot is located along the equatorial plane of the spin
axis ($\thetaS=90^\circ$). Even if we are viewing the WD from the
edge-on direction ($\thetao\approx 90^\circ$), the obliquity of the
spot is $\thetaS\approx 60^\circ$, implying that the spot region
  is significantly away from the polar regions of the spin axis of the
  rotating WD.

We conclude that ZTF J190132.9+145808.7 has a relatively large dark
spot. As inferred from the phase-resolved
spectrum~\citep{Caiazzo2021}, the dark spot is likely to be explained
in terms of the magnetic dichroism.  The spot structure may be
  also consistent with an off-centered dipole field as in the case of
  EUVE J0317-855~\citep{Ferrario1997}; the off-centered dipole field
  is an empirical configuration of the magnetic field, and assumes
  that its center differs from the stellar center by some offset along
  the dipole direction, while the surface brightness is still
  axisymmetric with respect to the dipole axis.  \citet{Ferrario1997}
  found that the off-centered dipole field explains the lightcurve and
  polarization of EUVE J0317-855. If ZTF J190132.9+145808.7 has a
  similar configuration, its obliquity angle between the rotation and
  magnetic axis should be very large ($\thetaS\approx 60^\circ$ or
  $90^\circ$), implying that the spot region, probably challenging a
  formation mechanism of such spots.

\section{Discussion and conclusion \label{sec:conclusion}}

We have presented a formulation of photometric variations of rotating
stars with an inhomogeneous surface intensity.  We have provided a set
of general expressions for the photometric lightcurve due to the
oblique stellar rotation under the assumption that the inhomogeneity
pattern is symmetric around an axis misaligned to the stellar spin
direction.  We have applied the formulation to two simple
inhomogeneity models that are supposed to approximate the surface of
WDs, and showed that they are able to reproduce the monochromatic
sinusoidal lightcurves observed for a majority of variable
WDs. Therefore, we conclude that the monochromatic sinusoidal
lightcurve is a fairly generic prediction for obliquely rotating
stars.

Furthermore, one of our surface models (the cap-model: \S
\ref{subsec:cap-model}) can model the departure from a purely
sinusoidal lightcurve that are observed for a fraction of WDs
including ZTF J190132.9+145808.7 \citep{Caiazzo2021}. The degree of
the parameter degeneracy has been discussed extensively in \S
\ref{subsec:application2ZTF}. Our precise model prediction for
  the lightcurve shape will be also useful to distinguish between the
  oblique rotation model and the pulsation signature together with
  their difference in the frequency power spectra. 

We note that the present method is not limited to white dwarfs alone,
and applicable to predict photometric variations of a wide class of
obliquely rotating stars, once their surface inhomogeneity
distribution is specified.  Indeed, our current models provide a very
useful parameterization that characterizes a physical model of the
surface distribution. Thus it is complementary to the inversion method
\citep[e.g.,][]{KF2010,KF2011,FK2012,Aizawa2020,Luger2021} to infer
the surface inhomogeneity distribution that does not require a priori
model but is challenging to reconstruct the surface model from the
limited signal-to-noise ratio of the observational data.

Even in purely sinusoidal lightcurves, the parameter degeneracy
  may be broken by combining spectro-polarimetric data.  Since the
  surface inhomogeneity is likely originated from the stellar magnetic
  field, the measurements of the longitudinal and surface components
  of the time-varying magnetic fields may reveal the orientation of
  the dipolar magnetic fields, i.e., $\thetao$ and $\thetaS$
  \citep{Landolfi1993} in a complementary manner.  Then we may break
  the parameter degeneracy from the joint fits to the
  spectro-polarimetric and photometric data. This methodology has been
  successfully applied to WD 1953-011 \citep{Valyavin2008} for
  instance, and will become applicable more precisely using our
  quantitative modeling of the photometric lightcurve.

According to our result, an appreciable non-sinusoidal signature
  appears on a lightcurve when both the obliquity angle $\thetaS$ and
  the stellar spin inclination angle relative to the observer
  $\thetao$ is larger than the spot size $\dts$. In this case, the
  spot becomes invisible from the observer for a certain rotation
  phase, and the photometric lightcurve in the phase exhibits
  plateaus. The plateaus appear either at the tops or the bottoms of
  the lightcurve when the spot has a lower or a higher intensity than
  the average of the stellar disk, respectively. The former
  corresponds to a dark spot ($I_s < I_0$) as in the case of ZTF
  J190132.9+145808.7. The latter cases with a bright spot ($I_s >
  I_0$), or a facula, could also occur for a strongly-magnetized
  fast-spinning WD, or a WD pulsar with a force-free magnetosphere, as
  inferred for neutron-star pulsars e.g., PSR
  J0030+0451~\citep{Bilous2019,Miller2019,Riley2019} and PSR
  J0740+6620~\citep{Riley2021, Miller2021}. The non-sinusoidal
  fraction of lightcurves of WDs in a certain temperature and mass
  range should be a useful statistical measure of their magnetospheric
  structure. We plan to perform systematic comparison of our
  lightcurve models against the photometric data obtained by Kepler,
  TESS, ZTF, and Tomo-e Gozen~\citep{Aizawa2022}, which will be
  reported elsewhere in due course.


\section*{Acknowledgements}
Y.S. thanks Kei Iida for his hospitality at Kochi University where the
manuscript of the present paper was completed.  Simulations and
analyses in this paper made use of a community-developed core Python
package for Astronomy, {\tt Astropy} \citep{price2018astropy}, and a
Python package for Kepler and TESS data analysis, {\tt Lightkurve}
\citep{lightkurve2018}. This work is supported by Grants-in Aid for
Scientific Research by the Japan Society for Promotion of Science,
Nos.18H012 (YS), 19H01947 (YS), 20K14512 (KF), 17K14248 (KK), and
18H04573 (KK).

\bibliographystyle{apj}
\bibliography{ref-suto}

\appendix

\section{Analytic derivation of integrals appearing
  in the expressions for photometric
  variations \label{app:derivation}}

Lightcurves for the cap-model without limb darkening are computed from
equation (\ref{eq:Lt-0-2b}), and rewritten as follows, depending on
the four different parameter ranges of $\gamma$ and $\dts$.
\begin{description}
\item[(a)] $0<\gamma<\pi/2$ and $\dts<\pi/2-\gamma$:
\begin{eqnarray}
\label{eq:Lt-a}
L_{\rm 0a}(t) &=&  \frac{\cos\gamma}{2 \overline{I}}
\left[I_s(1-\cos2\dts)+I_0(\cos2\dts+\cos2\gamma)\right]\cr
&& \qquad\qquad + \frac{2I_0\sin\gamma \cot^3\gamma}{\pi\overline{I}}
\int_{0}^{\pi} F_0(x;\gamma) dx
\end{eqnarray}
\item[(b)] $0<\gamma<\pi/2$ and $\pi/2-\gamma<\dts<\pi/2$:
\begin{eqnarray}
\label{eq:Lt-b}
L_{\rm 0b}(t) &=&  \frac{I_s\cos^3\gamma}{\overline{I}}
+\frac{2\sin\gamma\cot^3\gamma}{\pi\overline{I}}
\left({I_s}\int_{\varphics}^{\pi}F_0(x;\gamma) dx
+ {I_0}\int_{0}^{\varphics} F_0(x;\gamma)dx\right),
\end{eqnarray}
where $\cos\varphics \equiv -\cot\gamma\cot\dts$.
\item[(c)] $\pi/2<\gamma<\pi$ and $\dts<\gamma-\pi/2$:
\begin{eqnarray}
\label{eq:Lt-c}
L_{\rm 0c}(t) = \frac{I_0\cos^3\gamma}{\overline{I}}
+ \frac{2I_0\sin\gamma\cot^3\gamma}{\pi\overline{I}}
\int_{0}^{\pi} F_0(x;\gamma) dx .
\end{eqnarray}
\item[(d)] $\pi/2<\gamma<\pi$ and $\gamma-\pi/2 <\dts<\pi/2$:
\begin{eqnarray}
\label{eq:Lt-d}
L_{\rm 0d}(t) &=& \frac{I_0\cos^3\gamma}{\overline{I}}
+\frac{2\sin\gamma\cot^3\gamma}{\pi\overline{I}}
\left({I_s}\int_{0}^{\varphics} F_0(x;\gamma)dx
+{I_0}\int_{\varphics}^{\pi}F_0(x;\gamma) dx \right) .
\end{eqnarray}
\end{description}

Since equation (\ref{eq:Lt-c}) corresponds to a case where the spot is
invisible to the observer, it should reduce to $I_0/\overline{I}$.
This implies that the following definite integral of $F_0(x;\gamma)$
is given as
\begin{eqnarray}
  \label{eq:int-F0}
  \int_0^\pi F_0(x;\gamma) dx
  = \int_0^\pi
  \frac{\sin x(x\cos x-\sin x)}{(\cos^2 x +\cot^2\gamma)^2} dx
  = -\frac{\pi}{2} \frac{\sin^2\gamma}{|\cos\gamma|^3}
  (1-|\cos\gamma|^3) .
  \end{eqnarray}

Indeed, one can explicitly prove equation (\ref{eq:int-F0})
by decomposing it in three integrals as follows.
\begin{eqnarray}
\int_0^{\pi/2}
  \frac{x\sin x\cos x}{(\cos^2 x +\cot^2\gamma)^2} dx
  &&= \sin^4\gamma
\int_0^{\pi/2}
\frac{x\sin x\cos x}{(\cos^2\gamma\sin^2 x +\cos^2 x)^2} dx \cr
 &&= \frac{\pi}{4} \frac{\sin^4\gamma}{\cos^2\gamma (1+|\cos\gamma|)}.
  \end{eqnarray}
\begin{eqnarray}
\int_{\pi/2}^\pi
  \frac{x\sin x\cos x}{(\cos^2 x +\cot^2\gamma)^2} dx
  &&= -\pi
\int_0^{\pi/2} \frac{\sin x\cos x}{(\cos^2 x +\cot^2\gamma)^2} dx
+
\int_0^{\pi/2}
\frac{x\sin x\cos x}{(\cos^2 x +\cot^2\gamma)^2} dx \cr
&&= -\frac{\pi}{2} \frac{\sin^4\gamma}{\cos^2\gamma}
+ \frac{\pi}{4}
\frac{\sin^4\gamma}{\cos^2\gamma (1+|\cos\gamma|)}.
  \end{eqnarray}
\begin{eqnarray}
-\int_0^{\pi}
  \frac{\sin^2 x}{(\cos^2 x +\cot^2\gamma)^2} dx
  &&=
\int_0^{\pi}\frac{dx}{\cos^2 x +\cot^2\gamma}
- (1+\cot^2\gamma)
\int_0^{\pi}\frac{dx}{(\cos^2 x +\cot^2\gamma)^2}\cr
&&= \pi \frac{\sin^2\gamma}{|\cos\gamma|}
-\frac{\pi}{2} \sin^2\gamma
\frac{1+\cos^2\gamma}{|\cos\gamma|^3}.
  \end{eqnarray}
Summing up the above three equations leads to equation
(\ref{eq:int-F0}). Finally, substituting equation (\ref{eq:int-F0})
into equations (\ref{eq:Lt-a}) to (\ref{eq:Lt-d}) yields equations
(\ref{eq:L0a}) to (\ref{eq:L0d}) shown in \S \ref{subsec:cap-model}.

One can repeat the above argument that derives equation
(\ref{eq:int-F0}) for integrals of $F_1(x;\gamma)$ and
$F_2(x;\gamma)$. When we set $I(\theta)=I_0$, equations
(\ref{eq:Lt-1-2}) and (\ref{eq:Lt-2-2}) should be independent of
$\gamma$. So we evaluate their right-hand-sides for $\gamma=0$, and
find they reduce to $2I_0/3$ and $I_0/2$, respectively. Since the
result should hold for an arbitrary value of $\gamma$, we obtain the
following results:
\begin{eqnarray}
\label{eq:int-F1}
\int_{0}^{\pi} F_1(x;\gamma)dx=
\frac{2\pi}{3} \frac{\sin^5\gamma}{\cos^4\gamma}(5-3\sin ^{2} \gamma),
\end{eqnarray}
\begin{eqnarray}
\label{eq:int-F2}
\int_{0}^{\pi} F_2(x;\gamma)  dx
 = \pi\frac{\sin^2\gamma}{|\cos^5\gamma|}
 (21 |\cos^5\gamma|-15|\cos^7\gamma|-6).
\end{eqnarray}
These results are used in deriving equations (\ref{eq:Lt-1-pmodel}) and
(\ref{eq:Lt-2-pmodel}).

We also find that a long but straightforward algebra reveals
\begin{eqnarray}
 \int_{0}^{\pi} F_2(x;\gamma)
 \left(\frac{\cos x}{\sqrt{\cos^2 x+ \cot^2\gamma}}\right)^p dx
 = \frac{12\pi \sin^{p+6}\gamma \cos^4\gamma}{(p+2)(p+4)}
  [6+5(p+2)\cos^2\gamma],
\end{eqnarray}
if $p$ is an odd integer. Thus, we obtain an analytic expression as
the last line in equation (\ref{eq:J_2p}). We were not able to find an
analytic expression for equation (\ref{eq:J_1p}), which is evaluated
numerically in plotting Figure \ref{fig:Lt-LD-pmodel}.

\end{document}